\documentstyle[11pt,newpasp,twoside,epsf]{article}
\def\edcomment#1{\iffalse\marginpar{\raggedright\sl#1\/}\else\relax\fi}
\reversemarginpar

\begin{document}
\title{Disk Galaxies as Cosmological Benchmarks:\\
Cold Dark Matter versus Modified Newtonian Dynamics}

\author{Frank C. van den Bosch \& Julianne J. Dalcanton}
\affil{University of Washington, Box 351580, Seattle, WA 98195, USA}

\begin{abstract}
  We discuss a comparison of models for the formation of disk galaxies
  both in a Universe  dominated by cold dark  matter (CDM) and one  in
  which the force law is given  by modified Newtonian dynamics (MOND). 
  Our main aim is to address the claim made by McGaugh \& de Blok that
  CDM suffers from severe fine-tuning problems, which are circumvented
  under MOND.  As  we show, CDM  indeed requires some amount of tuning
  of the feedback efficiencies to obtain a Tully-Fisher relation (TFR)
  as  steep as observed.    However, that same  model is  in excellent
  agreement with   a   wide  variety  of  additional    observations.  
  Therefore,  the modest amount  of   feedback  needed should not   be
  regarded a fine-tuning problem.   Instead, its requirement should be
  considered a generic prediction for CDM, which  might be tested with
  future   observations   and  with    detailed modeling   of feedback
  processes.   We also show  that galaxy formation  in a MOND universe
  can not simultaneously   reproduce  the TFR   and the lack   of high
  surface  brightness dwarf galaxies.  We thus  conclude that CDM is a
  more viable theory for the formation of disk galaxies than MOND.
\end{abstract}

\section{Introduction}

Because   of their fairly   simple geometrical and dynamical structure
disk galaxies are the ideal benchmarks to test  our theories of galaxy
formation, gravity, and cosmology.  One notable characteristic of disk
galaxies  is   the flatness of  their   rotation curves, which implies
either that (i)  disks are embedded in a  massive dark matter halo, or
(ii) our theory of gravity is incorrect.   One theory build around the
latter option  is   Modified  Newtonian Dynamics  (MOND),   originally
proposed by  Milgrom  (1983a).  In MOND a  characteristic acceleration
$a_0$ is  assumed, such that for  accelerations  $a \ll a_0$  the true
acceleration  $a = \sqrt{a_0 a_N}$, with  $a_N$ the standard Newtonian
acceleration.  This implies flat rotation  curves, with an  asymptotic
value $V_{\infty}  = (G   M  a_0)^{1/4}$, without  the need   for dark
matter.

Here we examine how MOND fairs  in comparison to CDM  when it comes to
explaining  global properties of disk  galaxies  in the  context of  a
model for disk  galaxy  formation.  A  more detailed treatment  of the
models and results presented here can be found in van den Bosch (1998;
2000) and van den Bosch \& Dalcanton (2000).  Here we merely summarize
some of our results.

\section{The Tully-Fisher relation}

The zero-point, scatter, and  slope of the  TFR  of disk  galaxies ($L
\propto V_{\rm rot}^{\alpha}$) all  strongly depend on the photometric
band in which the TFR is  defined. In the  near-infrared, where one is
least sensitive to dust-extinction and  to uncertainties regarding the
ages and  metallicities   of stellar  populations,  one  finds $\alpha
\simeq 4$ (i.e., Verheijen 1997).

In the CDM scenario, one predicts a TFR of the form:
\begin{equation}
L \propto {\epsilon_{\rm gf} \over \Upsilon_d} V_{\rm vir}^{3}
\end{equation}
(see van den Bosch 2000 and references therein) with $V_{\rm vir}$ the
circular velocity at the virial radius, $\Upsilon_d$ the mass-to-light
ratio of the disk, and $\epsilon_{\rm  gf}$ a parameter that describes
what fraction of the  baryons inside the  virial radius resides in the
disk (as either  stars or  cold gas).   Thus,  $\epsilon_{\rm gf}$  is
related to the efficiencies of feedback and cooling.

In the case of MOND, direct application of the modified acceleration
results in a TFR of the form
\begin{equation}
L \propto {1 \over \Upsilon_d} V_{\infty}^{4},
\end{equation}
(see Milgrom 1983b).   Thus,  CDM and  MOND make distinct  predictions
regarding  the slope $\alpha$ of  the TFR. Furthermore, the prediction
for MOND is  in better agreement  with the near-infrared TFR.  For CDM
to yield  a TFR as  steep as observed  the process of galaxy formation
has to be  such that  $\epsilon_{\rm gf}  /  \Upsilon_d \propto V_{\rm
  vir}$.  Advocates of  MOND have used this  to argue against CDM,  as
they have claimed that  this requires extremely difficult  fine-tuning
(i.e,, McGaugh \& de Blok 1998a,b).
\begin{figure}
\plotone{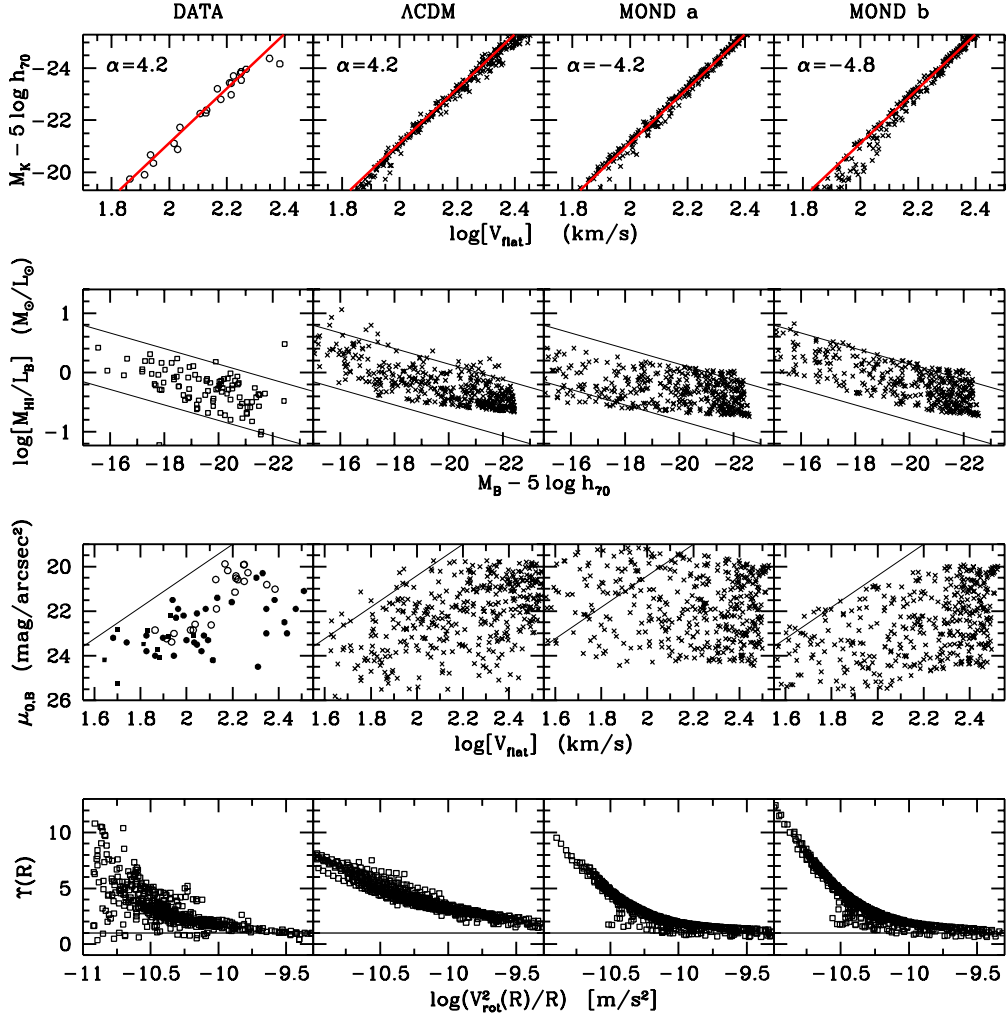}
\caption{A comparison of three models (one $\Lambda$CDM model and two
  MOND models) with various data on  disk galaxies (for the sources of
  the data see  van den Bosch   \& Dalcanton 2000). Upper  panels plot
  $K$-band  TFRs. The thick  solid lines indicate  the best-fit linear
  relation to  the data, and has  a slope of  $\alpha=4.2$.  Panels in
  the second    row plot the  gas mass   fraction  $M_{\rm HI}/L_B$ as
  function  of absolute $B$-band   magnitude. The  thin lines  have no
  physical meaning but are plotted to  facilitate a comparison between
  models and  data.  Panels  in  the third   row plot central  surface
  brightness versus  the amplitude of   the flat part  of the rotation
  curve.  Finally, the   panels  in the  lower row  plot  the enclosed
  mass-to-light  ratio  $\Upsilon(R)$  versus   the local acceleration
  $V_{\rm rot}^2(R)/R$.}
\end{figure}

In the upper panels of Figure~1 we plot  the $K$-band TFR for the data
as well as  for three of our  models: the $\Lambda$CDM model, and  two
MOND models ($a$ and $b$).  As can be seen, the $\Lambda$CDM TFR is as
steep as the data ($\alpha = 4.2$), which we accomplished by adjusting
the   feedback  efficiency.  Thus, indeed   some   amount of tuning is
required in  a  CDM Universe.  The  MOND  $a$  model,  which  has zero
feedback, also reveals  a TFR in   excellent agreement with the data.  
The  MOND $b$ model, however, in  which feedback is included, yields a
TFR that is too steep and with  too much scatter  at the low-mass end. 
The  reason  for the inclusion  of  the feedback in  this latter model
becomes apparent below.

\section{Other observational constraints}

The panels in the  second row of Figure~1 plot  the gas mass fractions
($M_{\rm HI}/L_B$) as function  of  the absolute magnitude.  The  data
reveals  a  systematic decrease of   $M_{\rm  HI}/L_B$ with increasing
magnitude, which is remarkably well reproduced  by the models and owes
in large part to  the stability related  threshold densities  for star
formation used in our models.

Panels in the  third row plot   the central surface  brightness of the
disk as function of $V_{\rm flat}$ (the  amplitude of the flat part of
the rotation curve).  The data   indicates an absence of high  surface
brightness (HSB) galaxies at  the low mass end  (indicated by the thin
solid line). This is nicely reproduced by  the $\Lambda$CDM model, and
owes to the  particular feedback model used.  In  the MOND  $a$ model,
however, no such  deficit of low-mass HSB  disks is present,  in clear
contradiction with  the data.  In  model $b$ we included feedback, for
which  we tuned the parameters  to reproduce  the data. Note, however,
that this model yields a TFR that is too steep.

Finally, in the lower panels of Figure~1 we plot the enclosed
mass-to-light ratio 
\begin{equation}
\Upsilon(R) = {R \, V_{\rm  rot}^2(R) \over G   \, M_{\rm disk}(R)},
\end{equation}
as  function of the  local   acceleration $V_{\rm rot}^2(R)/R$.   Here
$M_{\rm disk}$ is the `visible' mass of the disk (stars and cold gas),
and $R$ is the galactocentric  radius.  Each data point represents one
resolved measurement in the rotation  curve (RC) of  a disk galaxy. As
first  pointed out by  McGaugh (1998),  the  observed RCs of  galaxies
reveal a characteristic acceleration.   This is evident from  the fact
that the  data in  the lower-left panel  reveals very  little scatter. 
The presence  of  such a  characteristic  acceleration  is exactly the
`ansatz' of MOND,  and, not surprisingly,  the MOND models $a$ and $b$
nicely reproduce the data.  However, the $\Lambda$CDM model is also in
good  agreement with the  data.     Therefore,  the appearance  of   a
characteristic acceleration is not a unique prediction of MOND.

\section{Conclusions}

The  $\Lambda$CDM model requires tuning  of the feedback parameters to
yield a TFR as steep as observed.  However,  once this is achieved the
model reproduces  the gas mass fractions, the  absence of low mass HSB
galaxies, and reveals a characteristic acceleration  as observed.  All
the  particular problems  for CDM pointed   out by McGaugh \&  de Blok
(1998a)  are  in fact   solved by this   one simple  tuning.  The MOND
models, however,  are unable to  simultaneously  reproduce the TFR and
the absence of low-mass HSB disks.  Although we do not consider this a
strong proof against   MOND (as our  particular assumptions underlying
the model for galaxy formation  might not be  valid under MOND), we do
want  to    emphasize that our  results   largely  remove  the claimed
advantages of MOND over CDM.


\begin{references}

\reference McGaugh, S. S. 1998, preprint (astro-ph/9812327)

\reference McGaugh, S. S., \& de Blok, W. J. G. 1998a, \apj , 499, 41

\reference McGaugh, S. S., \& de Blok, W. J. G. 1998b, \apj , 499, 66

\reference Milgrom, M. 1983a, \apj , 270, 365

\reference Milgrom, M. 1983b, \apj , 270, 371

\reference van den Bosch, F. C. 1998, \apj , 507, 601

\reference van den Bosch, F. C. 2000, \apj , 530, 177

\reference van den Bosch, F. C., \& Dalcanton, J. J. 2000, \apj , 534,
146

\reference Verheijen, M. A. W. 1997, PhD Thesis, University of Groningen

\end{references}
\end{document}